\documentclass[graybox]{svmult}
\usepackage[numbers]{natbib}

\usepackage{helvet}         
\usepackage{courier}        
\usepackage{type1cm}        
\usepackage{makeidx}         
\usepackage{graphicx}        
\usepackage{multicol}        
\usepackage[bottom]{footmisc}
\renewcommand{\vec}[1]{{\mathbf{#1}}}

\newcommand{\beq}{\begin{eqnarray}}
\newcommand{\eeq}{\end{eqnarray}}

\renewcommand{\vec}[1]{{\mathbf{#1}}}

\usepackage{amsmath,amssymb,amsfonts}
\usepackage{braket}
\usepackage{upgreek}

\newcommand{\grad} {\nabla}
\newcommand{\lap} {\triangle}

\newcommand{\Ly}[1]{\lap_x #1 + \frac{a}{y} #1_y + #1_{yy}}

\newcommand{\R}{\mathbb R}

\renewcommand{\vec}[1]{\boldsymbol{#1}}
\newcommand{\p}{\ensuremath\uppi}

\def\a{\alpha}

\def\p{\partial}

\begin{document}

\title*{N\"other's Second Theorem as an Obstruction to Charge Quantization }
\author{Philip Phillips and Gabriele La Nave}
\institute{Philip Phillips \at Department of Physics and Institute of Condensed Matter Theory, University of Illinois at Urbana-Champaign, Urbana, Il. 61801-3080, \email{dimer@illinois.edu}
\and Gabriele La Nave \at Department of Mathematics, University of Illinois at Urbana-Champaign, Urbana, Il. 61801-3080 \email{lanave@illnois.edu}}
\maketitle

\abstract{  While it is a standard result in field theory that the scaling dimension of conserved currents and their associated gauge fields are determined strictly by dimensional analysis and hence cannot change under any amount of renormalization, it is also the case that the standard conservation laws for currents, $dJ=0$, remain unchanged in form  if any differential operator that commutes with the total exterior derivative, $[d,\hat Y]=0$, multiplies the current.   Such an operator, effectively changing the dimension of the current, increases the  allowable gauge transformations in electromagnetism and is at the heart of N\"other's  second theorem. We review here our recent work on one particular instance of this theorem, namely fractional electromagnetism and highlight the holographic dilaton models that exhibit such behavior and the physical consequences this theory has for charge quantization.  Namely, the standard electromagnetic gauge and the fractional counterpart cannot  both yield integer values of Planck's constant when they are integrated around a closed loop, thereby leading to a breakdown of charge quantization.   }

\section{Preliminaries}

Although N\"other\cite{nother} has two theorems, the second is little known but ultimately more important, as we will see.  The first theorem which sits at the foundation of gauge theories asserts that applying the gauge-invariant condition of electromagnetism $A_\mu\rightarrow A_\mu+\partial_\mu\Lambda$ to the Maxwell action
\beq
S=-\frac{1}{4}\int d^dx (F^2+J_\mu A^\mu)
\eeq
results in the conservation law
\beq
\partial_\mu J_\mu=0,
\eeq
with $F=dA$.  Because $\Lambda$ is a dimensionless function, $[A]=1$ and the current has fixed dimension $[J]=d-1$. Had we retained the dimensionful charge, we would have that $[qA]=1$.  Note the covariant derivative,
heuristically written as $D-iqA$, only fixes the dimension of the product
$[qA]=1$.  Hence, it is entirely possible to construct theories\cite{wise}
in which $q$ and $A$ have arbitrary dimensions without changing how the gauge group acts.  In what remains, we have set $q=1$ but our remarks apply to the dimensionful case as well.  The well known ambiguity (or ``improvement transformations''\cite{seiberg2019}) of the current,  namely that  the conservation laws remain fixed under shifting the current by a total derivative of the form, $J_0\rightarrow J_0+\partial_\mu X^\mu$ and $J_\mu\rightarrow J_\mu+\partial^0 X_\mu$, have no effect on the conserved charge nor the dimension of the current.  In fact as Gross\cite{gross,peskin} pointed out, because it is the action of the $U(1)$ group that ultimately fixes the dimension of the current through
\beq
\delta(x_0-y_0)[J_0(x),\phi(y)]=\delta \phi(y)\delta^d(x-y),
\eeq 
the dimension of the current, $[J_\mu]=d-1$, is sacrosanct unless one changes how the $U(1)$ group acts.  

This is the context\cite{nother} for NST.  N\"other\cite{nother} noticed that the form of the conservation law for the current is determined by the order of the derivative retained in the degeneracy condition for $A_\mu$.  In fact, there is no unique way of specifying this as can be seen from the following argument.  Consider the Maxwell action,
\beq\label{Maction}
S&=&\frac12\int \frac{d^dk}{2\pi^d} A_\mu(k)[k^2 \eta^{\mu\nu}-k^\mu k^\nu]A_\nu(k)\nonumber\\
&=&\frac12\int d^dk A_\mu(k)M^{\mu\nu}A_\nu(k).
\eeq
All gauge transformations arise as zero-eigenvalues of $M$.  For example,
\beq
M^{\mu\nu}k_\nu=0,
\eeq
which yields the standard gauge-invariant condition in electromagnetism because  $ik_\nu$ is just the Fourier transform of $\partial_\mu $.
The ambiguity that leads to NST comes from noticing that if $k_\nu$ is an eigenvector, then so is $f k_\nu$, where $f$ is a scalar. Whence, there are a whole family of eigenvectors,
\beq\label{mgen}
M_{\mu\nu}f k^\nu=0,
\eeq
that satisfy the zero eigenvalue condition, each characterizing a perfectly valid electromagnetism.   It is for this reason that N\"other\cite{nother} devoted the second half of her paper to the consequences of retaining all possible integer derivatives,
\beq
\label{Nothergen}
A_\mu\rightarrow A_\mu +\partial_\mu\Lambda +\partial_\mu\partial_\nu G^\nu+\cdots,
\eeq
 in the gauge-invariant condition for $A_\mu$ on the conservation laws for the current.   Stated succinctly, the second theorem finds that the full family of generators of $U(1)$ invariance determines the dimension of the current not just the linear derivative term $A_\mu\rightarrow A_\mu+\partial_\mu\Lambda$.   In general, the second theorem applies anytime there are either a collection of infinitesimal symmetries or one symmetry parametrized by an arbitrary number of functions as in Eq. (\ref{Nothergen}).  What N\"other\cite{nother} found is that the higher-order derivatives in the gauge-invariant condition add further constraints on the current.  They can even change the order of the current.  However, as long as only integer derivatives\cite{avery} are retained, the constraint equations yield no new content.  It is for this reason that N\"other's second theorem has garnered little interest. 
 
However, there is a generalization of Eq. (\ref{Nothergen}) that does yield non-trivial results. Consider the fact that the current conservation equation remains unchanged
if a differential operator $\hat Y$ exists such that $[d,{\hat Y}]=0$.  If such an operator exists then the conservation law becomes
\beq\label{Ycurrent}
\partial_\mu{\hat Y}J_\mu=\partial_\mu {\tilde J}_\mu=0,
\eeq
which the redefines the current to be ${\tilde J}={\hat Y}J$.   This is an ambiguity {\bf distinct} from the ``improvement transformations'' of the first theorem because  $\hat Y$ is linked to the  gauge symmetry.   We can construct $\hat Y$ directly from Eq. (\ref{mgen}).  Since $f k_\nu$ is the generator of the gauge symmetry, there are some constraints on $f$.  1) $f$ must be rotationally invariant.  2) $f$ cannot change the fact that $\Lambda$ is dimensionless; equivalently it cannot change the fact that $A$ is a 1-form.  3) $f$ must commute with the total exterior derivative; that is, $[f,k_\mu]=0$ just as $[d,{\hat Y}]=0$.  Hence, finding $f$ is equivalent to fixing $\hat Y$.  A form of $f$ that satisfies all of these constraints is $f\equiv f(k^2)$.   In momentum space, $k^2$ is simply the Fourier transform of the Laplacian, $-\Delta$.  As a result, the general form of $f(k^2)$ in real space is just the Laplacian
raised to an arbitrary power, and the generalization in Eq. (\ref{mgen}) implies that there are a multitude of possible electromagnetisms  (in vacuum) that are
 invariant under the transformation,
 \beq
 A_\mu\rightarrow A_\mu + f(k^2)ik_\mu\Lambda,
 \eeq
 or in real space,
  \beq
  A_\mu\rightarrow A_\mu + (-\Delta)^\gamma \partial_\mu\Lambda,
  \eeq
  resulting in $[A_\mu]=1+2[f]=\gamma$.  The definition of the fractional Laplacian we adopt here is due to Reisz:
\beq\label{reisz}
(-\Delta_x)^\gamma f(x)=C_{n,\gamma}\int_{\R^n}\frac{f(x)-f(\xi)}{\mid {x-\xi}\mid^{n+2\gamma}}\;d\xi
\eeq
for some constant $C_{n, \gamma}$.  Note rather than just depending on the information of $f(x)$ at a point, the fractional Laplacian requires information everywhere in ${\mathbb R}^n$. 
The standard Maxwell theory is just a special case in which $\gamma=1$.  In general, the theories that result for $\gamma\ne 1$ allow for the current to have an arbitrary dimension not necessarily $d-1$.  Identifying $\hat Y$ with the fractional Laplacian yields the conservation law
\beq \label{fractchargecons}
\partial^\mu (-\Delta)^{(\gamma-1)/2}J_\mu=0.
\eeq
\noindent
Conservation laws such as the one in Eq. (\ref{fractchargecons}) are in some sense more fundamental, as one can infer the standard ones from them but more importantly they can occur earlier\cite{gl1,csforms} in the hierarchy of conservation laws that stem from N\"other's first theorem.
This is the same conclusion reached from the degeneracy of the eigenvalue of Eq. (\ref{mgen}).   This consilience is not surprising because the degeneracy of the eigenvalue is another way of stating N\"other's second theorem.  That is, the current is not unique in gauge theories.  It is the lack of the uniqueness of the current that yields a breakdown of charge quantization.  As expected, this ambiguity shows up at the level of the Ward identities.  The current-current correlator for the photon
\beq
C^{ij}(k)  \propto (k^2)^{\gamma} \bigg( \eta^{ij} - \frac{k^ik^j}{k^2} \bigg)
\eeq
does not just satisfy  $k_\mu C^{\mu\nu}=0$ but also $k^{\gamma-1} k_\mu C^{\mu\nu}=0$. This translates into either $\partial_\mu C^{\mu\nu}=0$, the standard Ward identity, or
 \beq
 \partial_\mu (-\Delta)^{\frac{\gamma-1}{2}}C^{\mu\nu}=0
 \eeq
which illustrates beautifully the fact that the current conservation equation only specifies the current up to any operator that commutes with the total differential.  As we mentioned previously, this appears to be the first time this ambiguity has been linked to N\"other's Second Theorem.  

Because the fractional Laplacian is a non-local operator, the corresponding gauge theories are all non-local and offer a much broader formulation of electricity magnetism than previously thought possible.  All such anomalies can be understood as particular instances of N\"other's Second Theorem.  We will show how such theories arise from holographic bulk dilaton models\cite{g1,Gouteraux} and show that  Eq. (\ref{Ycurrent}) leads to a breakdown of charge quantization as can be seen from the fractional version of the Aharonov-Bohm effect\cite{ABnew,csforms}. 

\section{Charge Quantization}

Changing the dimension of the vector potential has profound consequences for the quantization of charge. This can be seen immediately because the integration of the gauge field around a closed loop
\beq\label{quant}
q\oint {\bf A}\cdot d\ell=h\mathbb Z
\eeq 
must be an integer multiple of Planck's constant, $h$.  This condition amounts to the integrability condition for the cohomology class of $qA$ to be an i{\it ntegral class}. Consequently, charge quantization is equivalent to the geometric requirement that the form $F_A=dA$ be indeed the curvature of a connection $D=d-qA$ on a $U(1)$ principal bundle $\mathcal P$.  It is on this fact that the Byers-Yang\cite{byersyang} theorem is based.  Clearly then when $[A]\ne 1$, the integral above is no longer dimensionless, leading to an inapplicability of the Byers-Yang theorem.  What is required in such cases is the construction of a new fictitious gauge field that does have the requisite dimension.  While the new gauge will preserve Eq. (\ref{quant}), the original one will not\cite{g1,ABnew}.  Consequently, if it is the fractional gauge field that describes the material in question, strictly speaking, charge is not quantized. That is, both gauges cannot preserve Eq. (\ref{quant}) simultaneously. Maxwell's equations amount to setting $f=1$ or $\gamma=1$.  As $f\ne 1$ is a perfectly valid electromagnetism, charge quantization is essentially a choice.  This is a physical consequence of N\"other's second theorem.

\section{Holographic Models with Fractional Gauge Transformations}

\begin{figure}
\center
	\includegraphics[scale=0.3]{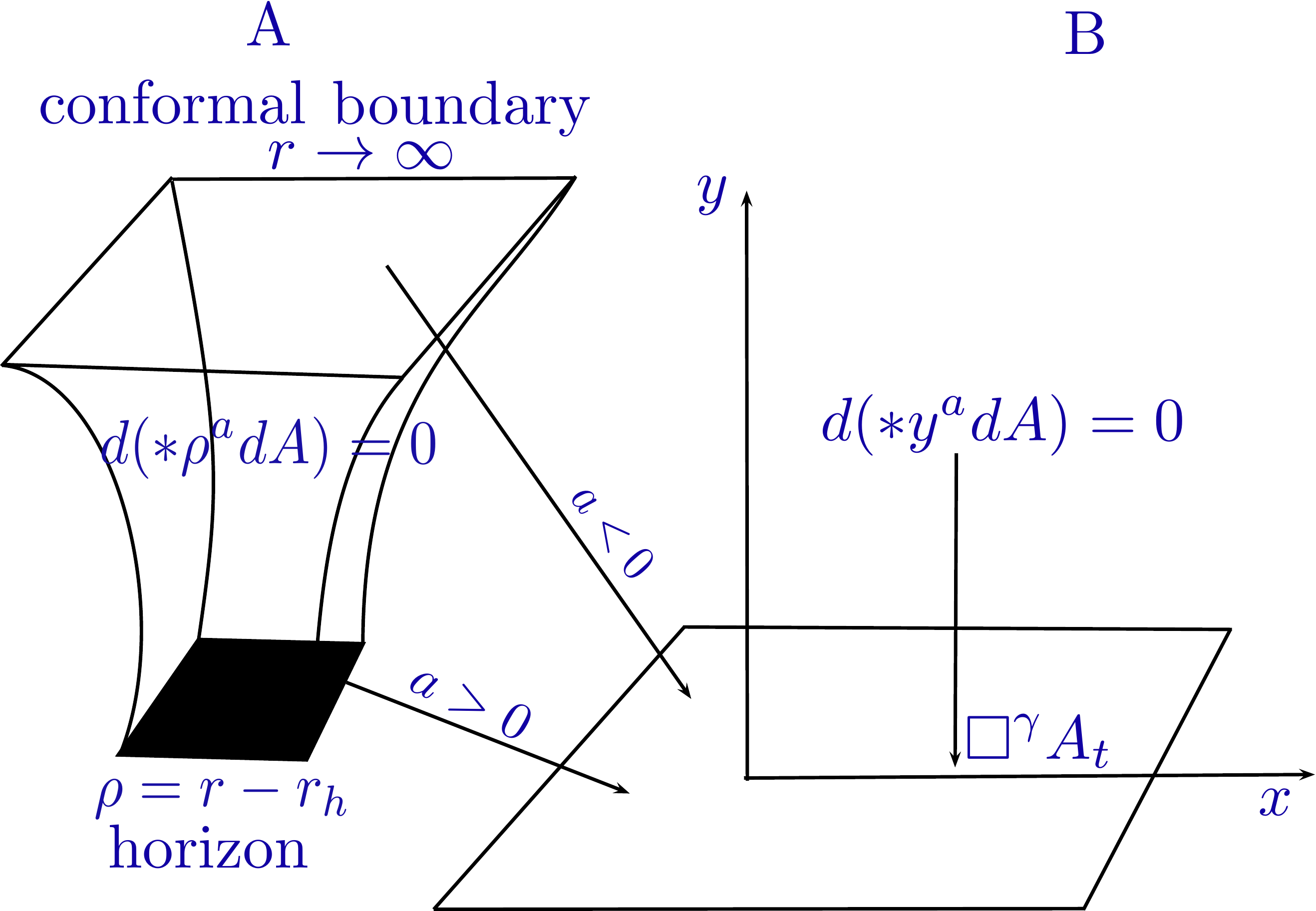} 
    \caption{A.) A depiction on an AdS spacetime with a conformal boundary at $r=\infty$ and a black hole horizon at $r=r_h$.  The Maxwell-dilaton action in the bulk has equations of motion of the form  $d(\ast \rho^a dA)=0$.  B.)  p-form generalization of the Caffarelli-Silvestre\cite{CS2007}-extension theorem. $A_t$ are the boundary (tangential) components of the bulk gauge field, $A$. For a dilaton action in $\mathcal R^n$ with the equations of motion $d(\ast y^a dA)=0$, the restriction of these equations of motion to the boundary yields the fractional Box operator where the exponent is given by $\gamma=(1-a)/2$.  Depending on the  sign of $a$, the bulk dilaton action either yields fractional Maxwell equations of motion at the conformal UV $(a<0)$ boundary or at the IR limit $(a>0)$ demarcated by the horizon radius, $r_h$.  } \label{pform}
\end{figure}

The  preliminaries lay play that within a model with local interactions and with $U(1)$ symmetry in tact, there is no way around Gross's\cite{gross} argument that the dimension of the gauge field and the current are fixed to $[A]=1$ and $[J]=d-1$, respectively.  However, N\"other's second theorem suggests that other possibilities exist.  Interestingly\cite{weinberg,wittensup}, superconductivity provides a simple counter example, in which  the current,
\beq\label{pipJ}
J_i=-\int \; K_{ij} (\vec x,\vec x')   \left( \vec A_j(\vec x')-\nabla _j'\phi(\vec x')\right) d^3x'
\eeq
has dimension $d-d_K-1$ and hence is a non-local function of the gauge field, $A(\vec x')$ as a result of the kernel $K_{ij}$ which arises from expanding the free energy around the minimum $\grad\phi-A=0$ with $\phi$ the $U(1)$ phase. The Pippard kernel\cite{pippard}, relevant to explaining the disorder dependence of the Meissner effect, amounts to a particular choice for $K_{ij}$.  Holographic constructions offer a possibility as a result of the extra dimension which allows for the boundary (either at the UV or the IR)  to have properties quite distinct from the bulk.  A distinct claim of dilatonic models of the form
\beq 
S=\int \mathrm{d}^{d+1}xdy\sqrt{-g}\left[\mathcal R-\frac{\partial\phi^2}2-\frac{Z(\phi)}4F^2+V(\phi)\right],\nonumber\\
\eeq
is that the boundary gauge field acquires an anomalous dimension that is determined solely by the asymptotic form of the action
\beq
S_{\rm Max}=\int dV_d dy(y^a F^2+\cdots),
\eeq
where $y$ is the radial coordinate in the anti-de Sitter spacetime.
That such models change the gauge structure at the boundary can be seen by interpreting the dilaton term $y^a$ as a running charge coupling $g(a)$ which depending on the exponent $a$ can yield a relevant interaction at either the UV or at the IR horizon.  In the standard holographic set-up\cite{Witten1998,klebanov}, the boundary lacks a global $U(1)$ structure only the bulk does where the gauge field acts as source for the boundary current. That is, the conformal boundary, which we denote by the zero of the radial coordinate, $y=0$, is not imbued by a local gauge structure in which $A(y=0,x)=A_\parallel +d\Lambda$.  More explicitly, once the boundary condition is set, $A(y=0,x)=A_\parallel$, the gauge degree of freedom is lost.  Of course, the gauge structure can be reinstated simply by changing the boundary conditions from Dirichlet to von Neumann.  Alternatively, the theory can have a non-trivial structure at the IR or at the horizon.  Theories valid at either the UV or the IR boundary can be constructed using the membrane paradigm\cite{thorne}.  In this case, this approach is particularly apropos as either the IR or UV limits are relevant depending on the value of $a$ as  can be seen from the equations of motion,
\beq\label{dilaton}
\nabla^\mu( \rho^aF_{\mu\nu})=0,
\eeq 
where we have introduced the radius $\rho=r-r_h$ which measures the distance from the horizon.  As depicted in Fig. (\ref{pform})
it is the IR limit which is relevant if $a>0$ and the UV in the opposite limit.  

To construct solve this boundary value problem, we appeal to a well known theorem in analysis.  In 2007, Caffarelli and Silvestre (CS)\cite{CS2007} proved that standard second-order elliptic differential equations in the upper half-plane in ${\mathbb R}_+^{n+1}$ reduce to one with the fractional Laplacian, $(-\Delta)^\gamma$, when one of the dimensions is eliminated to achieve ${\mathbb R}^n$.  For $\gamma=1/2$, the equation is non-degenerate and the well known reduction of the elliptic problem to that of Laplace's obtains. The precise statement of this highly influential theorem is as follows.  Let $f(x)$ be a smooth {\it bounded} function in ${\mathbb R}^n$ that we use to solve the extension problem,
\beq
g(x,y=0) &= &f(x) \nonumber\\
\Ly g &=& 0, \label{eq:withy}
\eeq
to yield a smooth {\it bounded} function, $g(x,y)$ in ${\mathbb R}^{n+1}_+$.
 $f(x)$ functions as the Dirichlet boundary condition of $g(x,y)$ at $y=0$.  These equations can be recast in degenerate elliptic form,
\beq\label{csext}
{\rm div}(y^a\nabla g)=0\quad{ \in}\,{\mathbb R}_+^{n+1},
\eeq
which CS proved has the property that 
\begin{equation}\label{caff-limit}
  \lim _{y\to 0^+} y^a \frac{\partial g }{\partial y} =C_{n,\gamma}\; {(-\lap)^\gamma f} 
\end{equation}
for some (explicit) constant $C_{n,\gamma}$ only depending on $d$ and $\gamma = \frac{1-a}{2}$
with $(-\lap)^\gamma$, the Reisz fractional Laplacian defined earlier.
That is, the fractional Laplacian serves as a Dirichlet to Neumann map for elliptic differential equations when the number of dimensions is reduced by one.  Consider a simple solution in which, $g(x,0)=b$, a constant, but also $g_x=0$.   This implies  that $g(y)=b+y^{1-a}h$ with $(1-a)>0$.  Imposing that the solution be bounded as $y\rightarrow\infty$ requires that $h=0$ leading to a vanishing of the LHS of Eq. (\ref{caff-limit}).  The RHS  also vanishes because $(-\Delta_x)^\gamma b=0$.  As a final note on the theorem, from the definition of the fractional Laplacian, it is clear that it is a non-local operator in the sense that it requires knowledge of the function everywhere in space for it to be computed at a single point.  In fact, it is explicitly an anti-local operator.  Anti locality of an operator $\hat T$ in a space $V(x)$ means that for any function $f(x)$, the only solution to $f(x)=0$ (for some $x\in V$) and $\hat T f(x)=0$ is $f(x)=0$ everywhere.  Fractional Laplacians naturally satisfy this property of anti-locality as can seen from their Fourier transform of Eq. (\ref{reisz}). 

Eq. (\ref{dilaton}) is highly reminiscent of Eq. (\ref{csext}) of the CS construction.  The only difference is that $g$ is a scalar in the CS-extension theorem and the gauge field is a 1-form.  Hence, the p-form generalization\cite{csforms} of the CS-extension theorem is precisely the tool we need to determine the gauge structure either at the conformal boundary or at the horizon.  The key ingredient in this proof is the fractional differential.  Because the Hodge Laplacian
\beq
\Delta=dd^\ast+d^\ast d:\Omega ^p (M)\to  \Omega ^p (M),
\eeq
does not change the order of a p-form, as it is a product of $d$ and $d^\ast$, it can be used to define the fractional differential
\beq \label{dgamma}
d_\gamma= d\Delta ^\frac{\gamma-1}{2}= \Delta ^\frac{\gamma-1}{2},\quad d_\gamma^\ast=d^\ast\Delta^\frac{\gamma-1}{2}.
\eeq
Since  $[d, \Delta ^b]=0$ for any power $b$, a key benefit of $d_\gamma$ is that the composition
\beq
(d_\gamma d_\gamma ^\ast+ d_\gamma ^\ast d_\gamma)\omega=\Delta^\gamma\omega
\eeq
offers a way of computing the action of fractional Laplacian on the differential form $\omega$.

These definitions allow an immediate construction of the p-form generalization of the CS-extension theorem.  for $\alpha\in\Omega^p$ and a bounded solution to the extension problem
\beq
\label{formsCS}
& d(y^ad^*\alpha)+d^*(y^a d\alpha)=0\in M\times \mathbb R_+\nonumber\\
& \alpha \mid _{\partial M} = \omega \text{ and } d^* \alpha \mid _{\partial M} =d_x^* \omega,
\eeq
then
\beq
\lim _{y\to 0} y^{a} {\rm i } _{\nu} d\alpha=C_{n,a} (\Delta) ^\gamma \omega,
\eeq
with $2\gamma= 1-a$ and where ${\rm i} _V \omega$ indicates the $(p-1)$-form determined by 
 ${\rm i} _V \omega (X_1,\cdots, X_{p-1})= \omega (X_1,\cdots, X_{p-1}, V)$,  $\nu = \frac{\partial}{\partial y}$, for some positive constant $C_{n,a}$.  This is the p-form generalization of the Caffarelli/Silvestre extension theorem.  It implies that the CS extension theorem on forms is the CS extension theorem on the components of the p-form.  The method of proof was simply component-by-component.  The succinct statement in terms of the components is easiest to formulate from the equations of motion
 \beq\label{formsCS2}
& {\rm div} (y^a \nabla \alpha _{i_1\cdots i_p})=0\in M\times \mathbb R_+\nonumber\\
& \left( \alpha _{i_1\cdots i_p}\right)\ \mid _{\partial M} = \omega _{i_1\cdots i_p}\ \text{ and } d^* \alpha \mid _{\partial M} =d_x^*\omega.
\eeq
Therefore, using the CS theorem, we have that 
\beq
\lim_{y\rightarrow 0} \; y^a \frac{\partial \alpha _{i_1\cdots i_p}} {\partial y} =C_{n,a}(-\Delta)^\a \ \omega _{i_1\cdots i_p},
\eeq
which proves that
\beq
\lim _{y\to 0} y^{a} {\rm i } _{\nu} d\alpha=(\Delta) ^a \omega,
\eeq
since by (elliptic) regularity of solutions to Eq. (\ref{formsCS})
\beq
\lim _{y\to 0} y^{a} \frac{\partial \alpha _{0 \ell _1, \cdots \ell _{p-1}} }{\partial x^{j_k}} =0.
\eeq

Applying this result to the dilaton equations of motion, Eq. (\ref{dilaton}) results in the fractional Maxwell equations 
\beq\label{fracmax}
\Delta^\gamma A_t=0.
\eeq
for the boundary components of the gauge field.  Since the only restriction is that $2\gamma=1-a$, this proof applies equally, with the use of the membrane paradigm\cite{thorne}, at the conformal boundary and the horizon.  Hence, even the dynamics in the IR (horizon) are governed by a fractional Maxwell action.  The curvature that generates these boundary equations of motion is 
\beq
F_\gamma = d_\gamma A= d\Delta ^\frac{\gamma-1}{2} A,
\eeq
with gauge-invariant condition,
\beq\label{eq:u1frac}
A \rightarrow A+ d_\gamma \Lambda,
\eeq
 where the fractional differential is as before in Eq. (\ref{dgamma}) which preserves the 1-form nature of the gauge field.
 This feature is guaranteed because by construction, the fractional Lagrangian cannot change the order of a form.
 As is evident,
 $[A_\mu]=\gamma$, rather than unity.  This gauge transformation is precisely of the form permitted by the preliminary considerations on N\"other's second
 theorem presented at the outset of this article and also consistent with the zero eigenvalue of the matrix $M$ in Eq. (\ref{Maction}).  
 
\section{N\"other's Second Theorem Revisited}
 
In order to determine how the fractional gauge acts, we first define the covariant derivative.  
To this end, we consider the ansatz,
 \beq\label{frac-covariant}
 D_{\alpha,\beta, A}\phi=(d+i\Box^\beta A)\Box^\alpha \phi,
\eeq
 with $\alpha$ and $\beta$ to be determined.  
The reason behind this ansatz for the covariant derivative is that we require the existence of a non-local transformation of the field $\phi$, the vector potential $A$ and the infinitesimal gauge group generator $\Lambda$ such that the covariant derivative transforms in the usual way $D_{\alpha,\beta, A}$ to the standard $D_{A'} \phi'=(d+iA') \phi'$ with the field redefinitions 
\beq\label{redef}
\phi'= \Box ^\alpha \phi \qquad A'=\Box ^\beta A.
\eeq
The Gauge action on $\phi'$ and $A'$ is thus the classical one
\beq
\phi'\to e^{i\Lambda '} \phi' \qquad A'\to A'+d\Lambda '
\eeq
and
\beq	
D_{A'} \left( e^{i\Lambda '}\phi'\right)=  e^{i\Lambda '}D_{A'+d\Lambda '} \phi'.
\eeq
Following the non-local transformations of Eq. (\ref{redef}), it is natural to suppose a field redefinition for the infinitesimal generators of the Gauge group as
\beq
\Lambda '=\Box^\delta \Lambda.
\eeq
Naturally, after such a change, there is only one way to define the Gauge group action,
\beq
e^{i\Lambda } \odot \phi= \Box^{-\alpha} \left( e^{i\Box^\delta \Lambda} \Box^\alpha \phi\right),
\eeq
to make $D_{\alpha,\beta, A}$ equivariant. 
The equivariant condition is then
\beq\label{equiv}
D_{\alpha,\beta, A}\left( e^{i\Lambda } \odot \phi\right)=  e^{i\Box^\delta \Lambda} D_{\alpha,\beta, A+d\Box^\delta \Lambda}\phi.
\eeq

We will define the curvature of $D_{\alpha,\beta, A}$ to be 
\beq\label{curv}
F_{\alpha,\beta, A}\phi= (d+i\Box^\beta A) D_{\alpha,\beta, A}\phi.
\eeq
This definition has the feature that it reduces to the curvature $F_{A'}$ after the transformations in Eq. (\ref{redef}).  In fact, it also reduces to the curvature $F_{\Box^\beta A}$ after the mere change of fields $\phi \to \phi'$.
At this point, we have not fixed the values of $\alpha$, $\beta$ and $\delta$. There are three natural conditions which we impose that will determine uniquely their values (hence the covariant derivative)  and the nature of the Gauge group action at the same time:
\begin{enumerate}
\item $D_{\alpha,\beta, A}\phi$ restricts to the fractional differential $d_\gamma \phi$ on functions when $A=0$.
\item The Gauge group action on connection fields must be $A\to A+d_\gamma \Lambda$.
\item The curvature $F_{\alpha,\beta, A}=i d_\gamma A$.
\end{enumerate}
The restriction that the covariant derivative reduce to the fractional differential, $d_\gamma$, when the fields are functions (Condition 1) imposes that $\alpha = \frac{\gamma-1}{2}$.
Next, we use Condition 2 to determine the value $\delta$. A quick read of Eq. (\ref{equiv}) will convince the reader that the Gauge transformation sends $A$ to $A+d\Box^\delta \Lambda$. Therefore, in order for condition 2 to hold, we require that $\delta =\frac{\gamma-1}{2}$.
Finally, in order to satisfy Condition 3, we make explicit the formula in Eq. (\ref{curv}).
\beq
F_{\alpha,\beta, A}\phi&=& (d+i\Box^\beta A) (d+i\Box^\beta A)\Box ^\alpha\phi\nonumber\\
&=& dd\Box^\alpha \phi+ i d\left( \Box ^\beta A\Box ^\alpha \phi\right) + i \Box ^\beta A d\Box ^\alpha \phi- \Box^\beta A\wedge  \Box^\beta A \Box^\alpha \phi\nonumber\\
&=&id(\Box ^\beta A) \phi-  i \Box ^\beta A d\Box ^\alpha \phi+ i \Box ^\beta A d\Box ^\alpha \phi= id(\Box ^\beta A) \phi.
\eeq
Therefore for Condition 3 to hold, we require that $\beta= \frac{\gamma-1}{2}$.
Summarizing, we have
\beq 
D_{\gamma, A} \phi &=& (d+i\Box^  \frac{\gamma-1}{2}A) \Box^  \frac{\gamma-1}{2}\phi\nonumber\\
e^{i\Lambda} \odot \phi&=& \Box^  \frac{1-\gamma}{2}\left( e^{i \Box^  \frac{\gamma-1}{2} \Lambda} \Box^  \frac{\gamma-1}{2}\phi\right)\nonumber\\
F_{\gamma, A} &=& (d+i\Box^  \frac{\gamma-1}{2}A) D_{\gamma, A} \phi,
\eeq
and the equivariance condition is
\beq
D_{\gamma, A} \left( e^{i\Lambda } \odot \phi\right)=  e^{i\Box^\frac{\gamma-1}{2}  \Lambda} D_{\alpha,\beta, A+d_\gamma \Lambda}\phi.
\eeq

We can now put N\"other's second theorem in this context of the redefined fields $A^\prime$ and $\phi^\prime$.  What we will show is that the standard version of N\"other's second theorem can be applied straightforwardly to a gauge action with $A^\prime$ and $\phi^\prime$ which can then be translated back to its non-local counterpart in terms of $A$ and $\phi$.  Given a schematic action for some field $\bf \Phi(x)$,
\beq
S(\Phi)= \int dx L(x, \bf\Phi(x)) ,
\eeq
we consider the infinitesimal action of Lie algebras and infinitesimal generators represented by vector fields
\beq
X = (D_IQ^\alpha)\frac{\p}{\partial \Phi_{,I}^\alpha},
\eeq
with $D_I= D_1^{i_1}\cdots D_p ^{i_p}$ for any multi-index $I=(i_1,\cdots , i_p)$, $ \Phi(x)^\alpha_{,I}=\frac{\p}{\p x^{i_1}}\cdots\frac{\p}{\p x^{i_p}}\Phi^\alpha$, and $D_m = \frac{\partial }{\partial x_m} + \sum _{\alpha, I} \frac{\partial \Phi_{,I} ^\alpha}{\partial x_m} \frac{\partial }{\partial \Phi_{,I} ^\alpha}$
by solving the (linearized) symmetry conditions
\beq
X(E_\alpha(L))=0,
\eeq
where $E_\alpha(L)$ are the Euler-Lagrange operators
\beq
E_\alpha(L) = (-1)^{|I|} D_IL
\eeq
and $|I|= \sum _m i_m$. The solution, $Q(\Phi)= (Q_1, \cdots Q_q)$ is called the characteristic of the symmetries generated by $X$.
N\"other's second theorem can now be stated as follows.
\begin{theorem}
The action $S$ admits an infinite dimensional group of symmetries with characteristics $Q(\Phi,B)$ that depend on arbitrary functions $B$ if and only if there exist differential operators $P_i$
such that
\beq\label{NSTI}
\sum _i P_i E_i(L)=0.
\eeq
\end{theorem}
In this language, the content of the second theorem is that $P_i$ are determined by the vector field $X$ and the statement is that there are infinitely many characteristics, that is, charges, if the sum preserves the total symmetry of the Euler-Lagrange equations.

We apply this discussion and N\"other's second theorem to the Lagrangian 
\beq
L '= \frac{1}{4} F'_{\mu \nu}F'^{\mu \nu} + (D_A')_\mu \phi '(D_A')^\mu \phi' + m^2 \phi'\phi'^*
\eeq
of the redefined fields $A'$, $\phi'$ defined in the previous section.
The Euler-Lagrange equations have components
\beq\label{EL}
\begin{aligned}
&E_{\phi'}(L)\equiv -\left( (D_A')_\mu (D_A')^\mu \phi' \right)^*+  m^2 \phi'^*=0\\
&E_{\phi'^*}(L)\equiv - (D_A')_\mu (D_A')^\mu \phi' +  m^2 \phi'=0\\
&E_{A_\mu} (L) \equiv i\phi' \left( (D_A')_\mu  \phi' \right)^*-i\phi'^* (D_A')_\mu \phi '+ \eta _{\mu \alpha} F^{\alpha \beta} _{,\beta}=0.
\end{aligned}\eeq
Amongst the variational symmetries, one finds the Gauge symmetries
\beq \phi'\to e^{i\Lambda '} \phi'; \qquad A' \to A' + d\Lambda '
\eeq
The generalized characteristics of the Gauge symmetries in components define an infinite set of charges
\beq Q^{\phi'} = -i \phi' B\qquad Q^{\phi'^*} = i (\phi')^* B\qquad Q^\mu= \eta ^{\mu \nu} B_{,\nu},
\eeq
for some arbitrary real function $B$.
The differential identity, Eq. (\ref{NSTI}), in N\"other's theorem is now
\beq
-i \phi' E_{\phi'}(L)+ i \phi'^* E_{\phi'^*}(L')- D_\alpha\left(   \eta ^{\mu \nu} E_\nu (L') \right)=0,
\eeq
where $D_\alpha$ is the total derivative. The key here is that the charges are arbitrary but yield nothing new\cite{avery} in the classical theory where only integer derivatives are present.
Clearly this is an operator equation as one can see by carrying out the calculations for $\int \mathcal{ D} \phi \mathcal {D} A e^{- S(\phi, A)}$ with arbitrary insertions.

When going back to the fields $A$, $\phi$ and the infinitesimal Gauge parameter $\Lambda$, we find that there is now a new non-trivial relation which gives rise to the action $A\to A +d_\gamma \Lambda$ and a new charge $Q= \int j_\gamma$ which did not exist in the theory corresponding to the action $S'$ (i.e., the classical Maxwell's equations). Effectively, one can see the fractional Maxwell equations as emergent from imposing the symmetry to be generated by the non-local action $e^{i\Box^\alpha \Lambda}$ for some $\alpha$.

\section{Aharonov-Bohm and Charge Quantization}

The inherent problem the degree of freedom $f$ (see Eq. (\ref{mgen})) introduces into electro-magnetism is that the multiplicity of gauge fields that are related by the fractional Laplacian, $A$ and $A^\prime$, each satisfy
\beq
 \int_\Sigma d_\gamma A=\oint_{\partial\Sigma} A^\prime,
 \eeq
 with $A^\prime=\Delta^{\frac{(\gamma-1)}{2}} A$.  As pointed out previously,   although this equality follows from Stokes' theorem, the result does not seem to have the units to be a {\it quantizable} flux.  That is, it is not simply an integer $\times hc/e$.  The implication is then that the charge depends on the scale.  In fact, because $[d,\Box^\gamma]=0$, the equations of motion can be rewritten as
 \beq\label{boxj}
 \Box^{\frac{\gamma-1}{2}}d(\star d\Box^{\frac{\gamma-1}{2}}A)=\star J.
 \eeq
 The current that emerges when Eq. (\ref{boxj}) is invertible has the equations of motion,
 \beq
 d(\star d \tilde{A})=\star \Box^{\frac{1-\gamma}{2}} J\equiv \star j.     
 \eeq
Similarly, the classical electromagnetic gauge $a \equiv \Box^{\frac{1-\gamma}{2}}A\equiv \Box^{1-\gamma} A^\prime$, hence having unit dimension, obeys the equations of motion
\beq \label{Max-a}  d(\star d a)= \Box ^{1-\gamma} j= \Box ^{ \frac{3}{2}(1-\gamma)}J.
 \eeq
Each of these choices for the gauge field defining different currents are all equally valid descriptions of nature.  The problem is that they are not all quantizable simultaneously.  For example, we have shown\cite{gl1} that 
\beq \label{alternative}
{\mathrm Norm}\left( \int_\ell  A^\prime\right)=\frac {\int_\ell A}{\Gamma(s+1)},
 \eeq
 with  $s= \frac{1-\gamma}{2}$, provided $\gamma<1$, and \beq\int_\ell  A^\prime=0\eeq if $\gamma >1$.  Hence, the line integral  $A$ or $ A^\prime$ cannot both yield integer values, the basic requirement for quantization.  Similarly, 
 \beq  {\rm Norm}\left( \int_\ell a \right)=\frac {\int_\ell A}{\Gamma(s+1)} 
 \eeq
with $s= \frac{1-\gamma}{2}$, when $\gamma >1$ and
\beq 
\int _\ell a =0
\eeq
when $\gamma <1$.  

\begin{figure}
\center
	\includegraphics[scale=0.5]{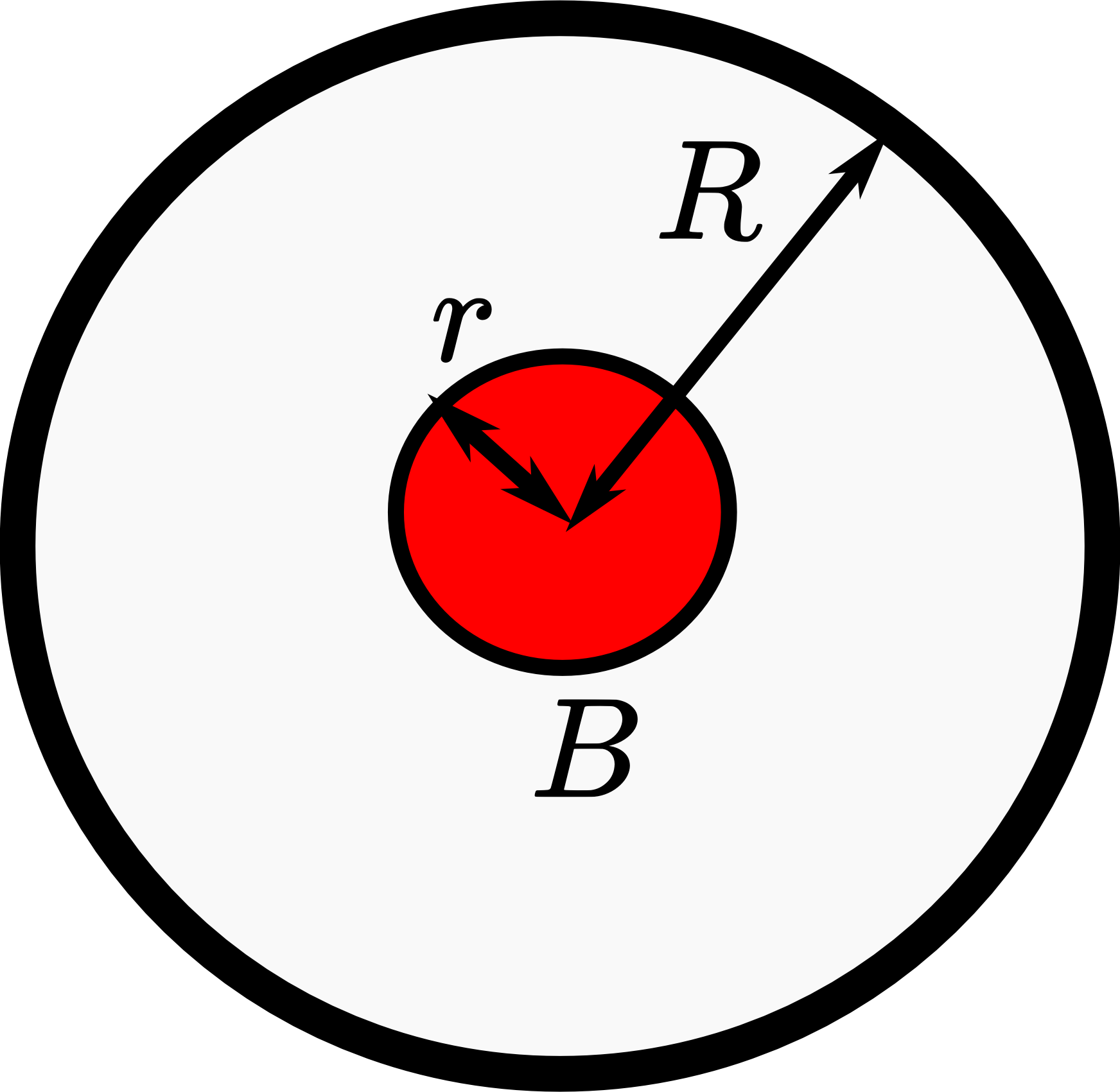} 
    \caption{Disk geometry for the Aharonov-Bohm phase\cite{ABnew}.  The fractional magnetic field pierces the disk in a small region of radius, $r$. }  \label{disk}
\end{figure}

All of this is a consequence of N\"other's second theorem: ambiguity in the gauge transformation leads to a breakdown of the standard charge
 quantization rules.  
What is the convention then for choosing the value of $\gamma$?  The answer is material dependent.  If either $A$ or $\tilde A^\prime$ are the physical gauge fields then the corresponding electric and magnetic fields in the material are indeed fractional. That is, each has an anomalous dimension.  Consequently, the flux enclosed in a disk of radius $r$ is no longer $\pi r^2 B$ simply because $[B]\ne 2$ and hence a failure of the key ingredient of the Byers-Yang theorem\cite{byersyang}.  The Aharonov-Bohm phase in this case for the disk geometry shown in Fig. (\ref{disk}) must be constructed by constructing using the fictitious gauge $a \equiv \Box^{\frac{1-\gamma}{2}}A\equiv \Box^{1-\gamma} A^\prime$ so that the correct dimensions are engineered in the usual covariant derivative $d-iqa$.  The result for the phase when $a$ is integrated around a closed loop \begin{eqnarray} \label{eq:phase_circle}
\Delta \phi_{\rm D} = \frac{e}{\hbar}\pi r^2 {_\alpha}B R^{2\alpha-2} \left(\frac{2^{2-2\alpha}\Gamma(2-\alpha)}{\Gamma(\alpha)} {_2}F_1(1-\alpha,2-\alpha,2;\frac{r^2}{R^2})\right).
\end{eqnarray}
involves the standard result, $\pi r^2 B$ multiplied by a quantity that depends on the total outer radius of the sample such that the total quantity is dimensionless.  Here  $_2F_1(a,b;c;z)$ is a hypergeometric function and the terms in the parenthesis reduce to unity in the limit $\alpha \rightarrow 1$.  This is the key experimental prediction of the fractional formulation of electricity and magnetism:  the  flux depends on the outer radius.  This stems from the non-local nature of the underlying theory and is the key signature that charge is no longer quantized in that is determined by a topological integral.

\section{Concluding Remarks}

In actuality, the ambiguity in defining the redundancy condition for the gauge field, Eq. (\ref{mgen}), ultimately leads to a landscape problem for charge quantization.  This is the physical import of N\"other's Second Theorem and the guiding mathematical idea behind our work on fractional electromagnetism\cite{csforms,gl1,ABnew}.   There is no easy fix here.  Each choice for $\gamma$ defines a valid vacuum theory of electromagnetism.  Ultimately it is a materials problem whether or not the fractional or standard gauge describe the interaction of matter with radiation.  In this sense, charge is ultimately emergent.

\begin{acknowledgement}
We thank K. Limtragool for a collaboration on the Aharonov-Bohm effect and E. Witten and S. Avery for insightful remarks and DMR19-19143 for partial support. 
\end{acknowledgement}
\bibliographystyle{apsrmp}

\end{document}